\documentclass[journal,comsoc]{IEEEtran}
\usepackage[T1]{fontenc}
\usepackage{color}
\usepackage{graphicx}
\usepackage[caption=false,font=footnotesize]{subfig}
\usepackage{fixltx2e}
\usepackage{float}

\ifCLASSINFOpdf
\else
\fi
\usepackage{amsmath}
\usepackage[cmintegrals]{newtxmath}
\begin{document}
%

\title{Adiabatic Floquet-Wave Solutions of Temporally Modulated Anisotropic Leaky-Wave Holograms}


%
%
%

\author{Amrollah~Amini,
        Homayoon~Oraizi,~\IEEEmembership{Life~Senior,~IEEE,}, Mostafa~Movahediqomi,
        and~Mohammad Moein~Moeini
\thanks{A. Amini and H. Oraizi are with the School
of Electrical Engineering, Iran University of Science and Technology, Tehran 1684613114,
Iran, e-mail: (h\_oraizi@iust.ac.ir).}
\thanks{M. Movahediqomi is with the Department of
 Electronics and Nanoengineering, School of Electrical Engineering, Aalto University, 02150 Espoo, Finland.}
\thanks{M. M. Moeini is with the Department of Electrical and Computer Engineering, Wayne State University, Detroit, MI 48202, USA.}
}

%
%

\markboth{}%
{Shell \MakeLowercase{\textit{et al.}}: Bare Demo of IEEEtran.cls for IEEE Communications Society Journals}
%



\maketitle

\begin{abstract}
  In this paper, the aperture field estimation technique and generalized adiabatic Floquet-wave expansion method is developed for two-dimensional temporally modulated anisotropic holograms.
  The determinant equations for propagation eigenstates in the presence of space-time modulation are obtained, and the behaviors of the corresponding dispersion curves are investigated. 
 Injecting temporal modulation causes asymmetrical displacement of the dispersion curve, which is studied in detail in this paper.
  The consequence of temporal modulation is the Doppler-shift effect and nonreciprocal response observed in these structures. 
  Results show that its nonreciprocity can be controlled by varying the modulation frequency injected to the hologram.
  The proposed method  significantly expands the range of applications of leaky-wave holograms.
  The main advantage of leaky-type modulated metasurfaces over the transmissive/reflective counterparts is their non-protruding feeds making them suitable for integrated structures. As an example, a nonreciprocal circularly polarized hologram is designed. Tensorial impedance surface is exploited to achieve high polarization purity.
  
\end{abstract}


\IEEEpeerreviewmaketitle

\section{Introduction}
\IEEEPARstart{H}{aving} the ability to break tangential momentum constraints, spatially modulated metasurfaces can be exploited in realizing a wide range of microwave and optical devices, such as reflective/transmissive \cite{yu2011}-\cite{arbabi2015} and leaky-wave components \cite{fong2010}-\cite{bodehou2019}. However, these structures suffer from some restrictions, such as linearity and reciprocity. Therefore, for the achievement of novel physical behavior, it is necessary to move beyond the conventional structures towards the next generation devices, namely dynamic metasurfaces \cite{shaltout2019}-\cite{cardin2020}. Imparting of temporal modulation to space-gradient metasurfaces removes the energy conservation constraints opening up unprecedented  opportunities to realize some physical phenomena, such as nonreciprocity unachievable by magnet-free static metasurfaces \cite{correas2015}-\cite{hu2021}.
Furthermore, by leveraging the temporal modulation, we may dispense with the need for traditional radio-frequency (RF) circuits and devices, like mixers, duplexers, power amplifiers, and filters in communication systems \cite{taravati2017_TAP}.
In addition to the scattering-type metasurfaces, the leaky-wave counterparts with time modulation have been studied as a radiator or receiver in order to transform the surface wave to the space wave or vice versa. Reference \cite{galiffi2020} proposed a new temporal structure to manipulate the surface waves and alleviating the high-cost, and high-loss features of the advanced nano-structures, but providing a precise control over the near-field wave performance to get more functionalities in a fast-switchable fashion.
The prominent advantage of leaky-wave metasurfaces is their integrability due to the planar feeding networks in contrast with the reflective/transmissive metasurfaces being excited by protruding feeding systems \cite{taravati2017_TAP}-\cite{amini2022}.
This unique feature makes the leaky type metasurfaces more suitable for assembling with other communication systems and removing some adverse effects concomitant with the diffraction phenomena such as shadow effects \cite{nayeri2014}.
On the other hand, a few works on dynamic metasurfaces have been reported to synthesize leaky-wave metasurfaces.

An effective approach in the design and synthesis of leaky-wave metasurfaces is the holographic technique, which comes from optics as a synthesis method \cite{gabor1948}, \cite{hariharan1996}. Reference \cite{fong2010} proposed the application of holographic technique in the microwave systems to achieve high gain leaky-wave radiators.
Based on this technique an interferogram (as a hologram) is constructed by the interference between a reference wave and an object wave at the first step. Then, the hologram is illuminated by the reference wave to realize the desired far-field pattern. 
This approach facilitates the design process of a leaky-wave metasurface using a simple procedure. However, for controlling all aspects of the radiation by leaky-wave metasurfaces  (including linear momentum, spin angular momentum (SAM), and the vorticity states), the metasurfaces should be first designed analytically.
For example, aperture field estimation (AFE) technique \cite{minatti2015}, \cite{teniou2017}, Method of Moments (MoM) \cite{ovejero2015}, \cite{bodehou2019_mom}, and the Floquet-wave expansion method \cite{minatti2016_FO}, \cite{amini2021} are among the most efficient techniques for analyzing the leaky-wave metasurfaces.

In this work, we generalize the adiabatic Floquet-wave expansion method for the application of dynamic two-dimensional leaky-wave holograms based on the spatio-temporally modulated boundary condition \cite{cassedy1965}. In \cite{cassedy1965} the interaction between electromagnetic waves with spatio-temporally modulated surfaces in a travelling-wave sense were used for open waveguide structures. The author rendered a bound solution for such structures where the surface waves were confined to the dielectric waveguides and the region of the solution was restricted to the slow-wave region on the Brillouin diagram. In our work, we extend this method for two-dimensional leaky-wave holograms in the fast-wave region. 
Besides by exploiting the aperture field estimation (AFE) method, the desired aperture distribution is mapped to the optimal surface impedance throughout the proposed structure. By doing so, we benefit from a precise systematic method to analyze and synthesize a dynamic leaky-wave metasurface in which a full control on all the radiation characteristics is achieved. Finally, we achieve the nonreciprocity as well as Doppler like shift effect for the proposed prototype to validate the accuracy of the application of Floquet-wave expansion method  to the dynamic leaky-wave metasurfaces.
\section{Impedance boundary condition and Floquet-wave expansion}
\subsection{Temporally modulated impedance boundary condition}
\begin{figure}
  \centering
  \includegraphics[width=0.4\textwidth]{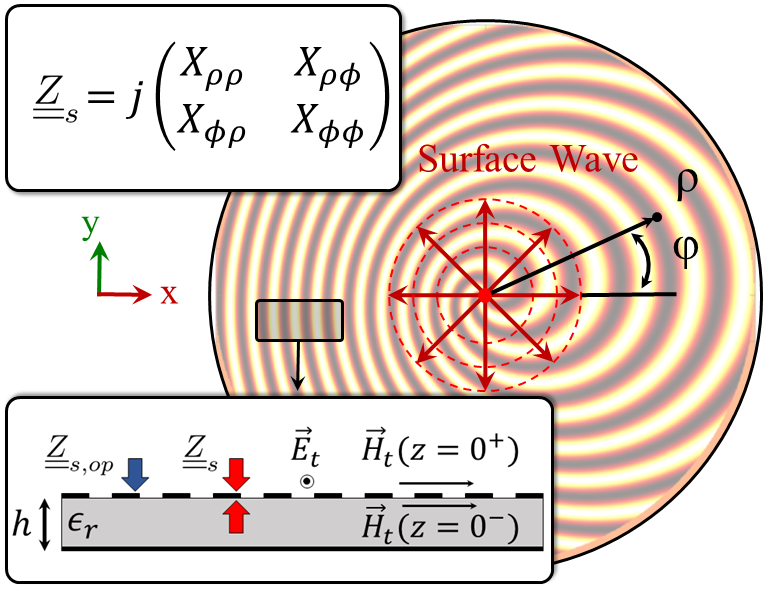}
  \caption{Conceptual schematic diagram of a leaky-wave metasurface. The inset depicts the opaque and transparent impedance boundary conditions. The surface wave launcher is located at the center of metasurface, which excites the cylindrical surface waves.}
  \label{fig:Fig1}
\end{figure}

In the general case, a leaky-wave metasurface can be described by an impedance boundary condition in which the average tangential electric field ($\vec{E}_t$) relates to the tangential magnetic field ($\vec{H}_t$) by defining a space-time variant surface impedance tensor as follows \cite{cassedy1965}:
\begin{equation}
\vec{E}_t(\vec{\rho},t)|_{z=0^+} = \underline{\underline{Z}}_{s,op}(\vec{\rho}, t) \cdot \hat{z}\times\vec{H}_t(\vec{\rho},t)|_{z=0^+}
\end{equation}
We define $\underline{\underline{Z}}_{s,op}$ as the opaque impedance since it consists of both impedance cladding and grounded dielectric. Another definition can be given just for the cladding by considering the difference of magnetic fields through the following boundary condition:
\begin{equation}
\vec{E}_t(\vec{\rho},t)|_{z=0^+} = \underline{\underline{Z}}_s(\vec{\rho}, t) \cdot \hat{z} \times (\vec{H}_t(\vec{\rho}, t)|_{z=0^+} - \vec{H}_t(\vec{\rho}, t)|_{z=0^-})
\label{eq:imp_trans}
\end{equation}
The former representation is named as the transparent impedance. In the absence of losses, $\underline{\underline{Z}}_{s, op}$ and $\underline{\underline{Z}}_s$ are anti-Hermitian tensors \cite{minatti2016}. Both tensors are functions of $\vec{\rho}$ and $t$. 
By embedding a surface wave launcher in the surface as the excitation source and properly modulating the boundary condition, the interaction between the surface wave and the impedance surface may generate a polarized leaky wave with arbitrary linear and angular momentum states.
Fig. \ref{fig:Fig1} shows the schematic diagram of a leaky-wave metasurface with modulated impedance boundary condition.
In this paper we utilize the transparent impedance throughout our analysis.

In order for the generalized Floquet-wave expansion to be applied, we assume that the tensorial surface impedance is a locally periodic function in space and time acting as a travelling wave. 
Generally, we may express the impedance tensor as follows:
\begin{equation}
\underline{\underline{Z}}_s(\vec{\rho}, t) = j( \hat{\rho}\hat{\rho}X_{\rho\rho} + (\hat{\rho}\hat{\phi} + \hat{\phi}\hat{\rho})X_{\rho\phi} + \hat{\phi}\hat{\phi}X_{\phi\phi})
\label{eq:Z_s}
\end{equation}
\begin{equation}
X_{\rho\rho} = \bar{X}_{\rho} [1 + m_\rho(\vec{\rho})  \cos (Ks(\vec{\rho}, t) + \Phi_\rho(\vec{\rho}))]
\label{eq:X_rho_rho}
\end{equation}
\begin{equation}
X_{\rho\phi} = \bar{X}_{\rho} [m_\phi(\vec{\rho})  \cos (Ks(\vec{\rho}, t) + \Phi_\phi(\vec{\rho}))]
\label{eq:X_rho_phi}
\end{equation}
\begin{equation}
X_{\phi\phi} = \bar{X}_{\phi} [1 - m_\rho(\vec{\rho})  \cos (Ks(\vec{\rho}, t) + \Phi_\rho(\vec{\rho}))]
\label{eq:X_phi_phi}
\end{equation}
where $\bar{X}_\rho$ and $\bar{X}_\phi$ are average values of $X_{\rho\rho}$ and $X_{\phi\phi}$, respectively. Note that, $Ks(\vec{\rho}, t)$ and $\Phi_{\rho, \phi}(\vec{\rho})$ are the rapidly- and slowly-varying parts of the modulation phase, respectively such that:
\begin{equation}
|\nabla_{\vec{\rho}}Ks(\vec{\rho}, t)| \gg |\nabla_{\vec{\rho}}\Phi_{\rho, \phi}(\vec{\rho})|
\end{equation}
where $\nabla_{\vec{\rho}} = \nabla - \hat{z}\hat{z}\cdot \nabla$ indicates the transverse gradient operator on $\vec{\rho}$.
The local periodicity of the impedance boundary condition can be defined in the $\rho$ direction as:
\begin{equation}
P = |\frac{2\pi}{\nabla_{\vec{\rho}} Ks(\vec{\rho}, t)\cdot \hat{\rho}}|
\label{eq:P}
\end{equation}
Meanwhile, the temporal periodicity can also be obtained as:
\begin{equation}
T=|\frac{2\pi}{\partial_tKs(\vec{\rho}, t)}|
\label{eq:T}
\end{equation}
where $\partial_t$ denotes the derivation on time.
Note that the linear and orbital angular momentums of radiated wave are determined by $Ks(\vec{\rho}, t)$ and $\Phi_{\rho, \phi}(\vec{\rho})$, respectively, which affect in the phase of surface impedance tensor elements. Besides, the spin angular momentum (polarization state) is assigned by the relationship between $m_\rho(\vec{\rho})$ and $m_\phi(\vec{\rho})$ \cite{amini2021}.
\subsection{Adiabatic Floquet-wave expansion}
We now expand the surface current by the adiabatic Floquet modes for spatio-temporally modulated metasurfaces. Note that this current expansion has been previously implemented by the Flat Optics for a pure-space modulated case \cite{minatti2016_FO}. According to the periodicity of surface impedance (Equations (\ref{eq:P}) and (\ref{eq:T})) the higher-order modes need to be considered for the surface current. So the surface current may be expressed in terms of $n$'th harmonic:
\begin{equation}
  \vec{J}\approx e^{j\omega t}\sum_{n=-\infty}^{\infty}(J_\rho^{(n)}\hat{\rho} + J_\phi^{(n)}\hat{\phi})e^{-jnKs(\vec{\rho}, t)}H_1^{(2)}(\tilde{k}_{sw}\rho)
  \label{eq:J}
  \end{equation}
  where $H_1^{(2)}$ is the first order Hankel function of second type, depicting the cylindrical wavefront 
  excited by the surface wave launcher at the center of structure (see Fig. \ref{fig:Fig1}). 
  The above equation may be simplified by the asymptotic form of the Hankel function:
  \begin{equation}
    \vec{J}\approx e^{j\omega t}\sum_{n=-\infty}^\infty \sqrt{\frac{2j}{\pi \tilde{k}_{sw}\rho}} (J_\rho^{(n)}\hat{\rho} + J_\phi^{(n)}\hat{\phi})e^{-j(nKs(\vec{\rho}, t) + \tilde{k}_{sw}\rho)}
    \label{eq:J_approx}
    \end{equation}
    Equation (\ref{eq:J_approx}) can be considered as a generalization of the Floquet-wave expansion proposed in \cite{cassedy1965}, which is valid for the two-dimensional modulated structures with anisotropic boundary conditions. 
    By applying the derivation of the phase of (\ref{eq:J_approx}) in terms of space and time, the spatial and temporal frequencies can be extracted, respectively:
    \begin{equation}
    \vec{k}^{(n)} = \nabla_{\vec{\rho}}[\tilde{k}_{sw}\rho + nKs(\vec{\rho}, t)]
    \label{eq:k_n}
    \end{equation}
    \begin{equation}
    \omega^{(n)} = \omega - n\partial_tKs(\vec{\rho}, t)
    \label{eq:omega_n}
    \end{equation}
    The Doppler-shift effect can be observed from (\ref{eq:omega_n}) due to the temporal variation in the modulation phase.
    Furthermore, the local wave-vector and the local leakage factor are obtained as follows:
    \begin{equation}
    \vec{\beta}^{(n)}=Re\{\nabla_{\vec{\rho}}[\tilde{k}_{sw}\rho + nKs(\vec{\rho}, t)]\}
    \end{equation}
    \begin{equation}
    \vec{\alpha} = -Im\{\nabla_{\vec{\rho}}\tilde{k}_{sw}\}
    \end{equation}
  \section{Estimating modulation parameters using aperture field estimation technique}
  Before analyzing the space-time modulated impedance boundary condition, in the following (Section \ref{subsection:pure_space}), we first study the synthesis process of pure-space modulated holograms. 
  Aperture field estimation method is used to synthesize a hologram generating circularly polarized wave with high polarization purity. Using this method, the modulation parameters (namely $\bar{Ks}(\vec{\rho})$, $m_{\rho, \phi}(\vec{\rho})$, and $\Phi_{\rho, \phi}(\vec{\rho})$) can be determined precisely.
Then we add the temporal variations to the modulation phase in the form of $Ks(\vec{\rho}, t) = \bar{Ks}(\vec{\rho}) - \omega_p t$, which will be discussed in Section  \ref{subsection:TV_hologram}.
  \subsection{Pure-space modulated hologram \label{subsection:pure_space}}
\begin{figure*}
  \centering
       \subfloat[\label{fig:delta_alpha}]
       {%
         \includegraphics[width=0.32\textwidth]{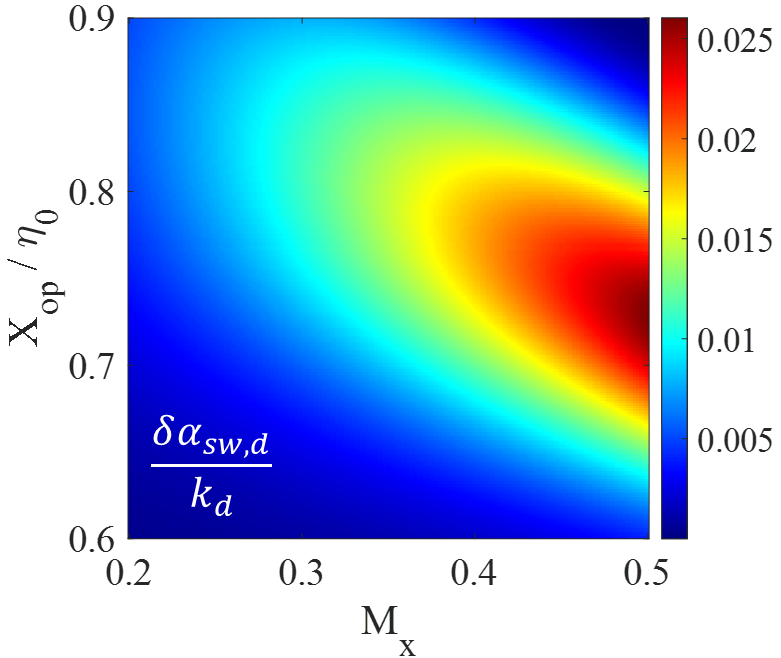}
       } 
      \subfloat[\label{fig:beta}]
      {
         \includegraphics[width=0.31\textwidth]{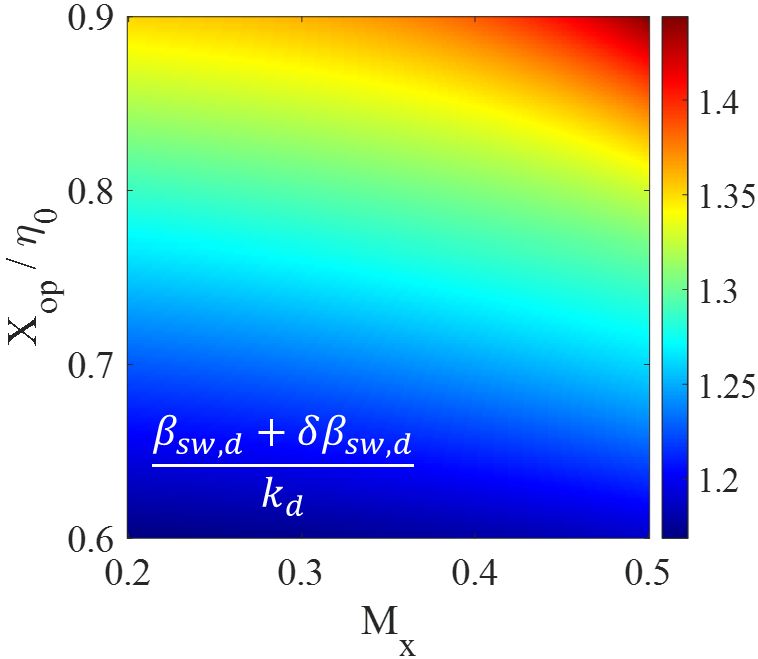}
      }
      \subfloat[\label{fig:error}]
      {
         \includegraphics[width=0.32\textwidth]{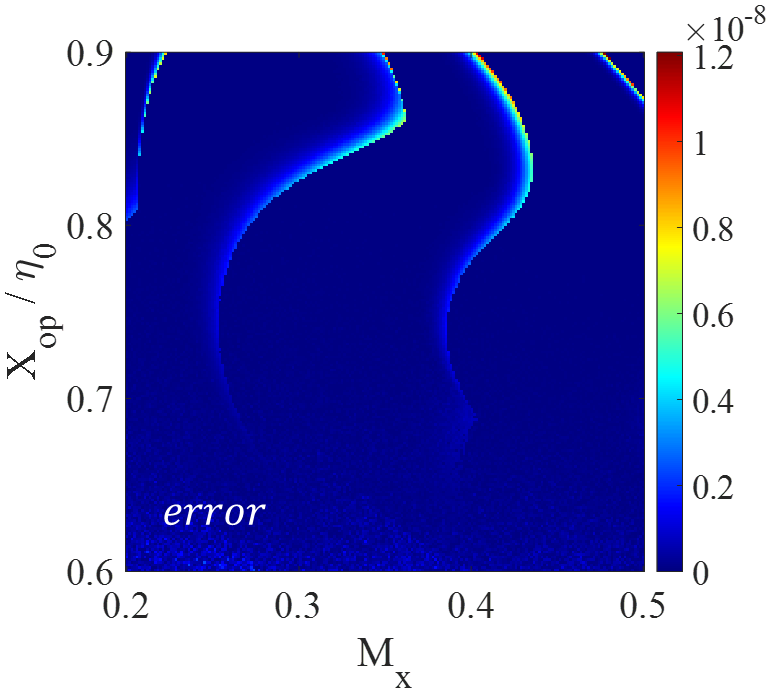}
      }
    \caption{Analytical results calculated for a pure-space modulated hologram versus modulation index and opaque reactance. (a) attenuation constant ($\delta \alpha_{sw}$). (b) phase constant ($\beta_{sw, d} + \delta \beta_{sw, d}$). (c) calculation error.}
    \label{fig:color_map_alpha_beta}
\end{figure*}

  An effective method of synthesizing static holographic antennas is the aperture field estimation technique originally proposed in \cite{minatti2015}. In this method the surface impedance distribution required to achieve the desired aperture field is accurately estimated. Its advantage is the simultaneous synthesis of the phase and amplitude of the aperture field, which is desirable for obtaining various shaped beams. 
  In the aperture field estimation method, the relationship between the aperture electric field ($\vec{E}_{ap}(\vec{\rho})$) and the surface impedance tensor is obtained as \cite{minatti2015}:
  \begin{equation}
  \underline{\underline{Z}}_s(\vec{\rho}).\hat{\rho} = jX_0[\hat{\rho} + 2Im\{\frac{\vec{E}_{ap}(\vec{\rho})}{-J_{sw}H_1^{(2)}(\tilde{k}_{sw,d}\rho)}\}]
  \label{eq:Zs_Eap}
  \end{equation}
  where $\tilde{k}_{sw, d}$ is the surface wave propagation number at the design frequency and may be evaluated as \cite{minatti2016_FO}:
  \begin{equation}
    \frac{\partial}{\partial \rho}(\tilde{k}_{sw,d} \rho) = k_{sw, d} = \beta_{sw, d} + \delta \beta_{sw,d} - j \delta \alpha_{sw,d}
  \end{equation}
  Note that, $\beta_{sw,d}$ is the propagation constant of cylindrical wavefront for unmodulated case, which has a small deviation $\delta \beta_{sw,d} - j \delta \alpha_{sw,d}$ in the presence of space modulation.
  In general case, the complex deviation $\delta \beta_{sw,d} - j \delta \alpha_{sw,d}$ is not independent of $\vec{\rho}$. Therefore, we have:
  \begin{equation}
    \tilde{k}_{sw,d} = \beta_{sw, d} + \delta \tilde{\beta}_{sw, d} - j \delta \tilde{\alpha}_{sw, d}
  \end{equation}
  where:
  \begin{equation}
    \delta \tilde{\beta}_{sw,d} = \frac{1}{\rho} \int_0^\rho \delta \beta_{sw,d} d\rho'
  \end{equation}
  \begin{equation}
    \delta \tilde{\alpha}_{sw,d} = \frac{1}{\rho} \int_0^\rho \delta \alpha_{sw,d} d\rho'
  \end{equation}
  To have a beam in the direction of $\theta = \theta_0$ and $\phi = \phi_0$, we define the x and y components of aperture field as follows \cite{amini2021}:
  \begin{equation}
    \begin{split}
  E_{ap, x}(\vec{\rho}) = M_x(\vec{\rho}) \frac{J_{sw}}{\sqrt{2\pi\rho|\tilde{k}_{sw,d}|}} e^{-\delta \tilde{\alpha}_{sw,d} \rho} \times\\
   e^{-j(k_d\rho \sin\theta_0 \cos (\phi - \phi_0) + l\phi)}
    \end{split}
  \label{eq:Eap_x}
  \end{equation}
  \begin{equation}
    \begin{split}
  E_{ap, y}(\vec{\rho}) = M_y(\vec{\rho}) \frac{J_{sw}}{\sqrt{2\pi\rho|\tilde{k}_{sw,d}|}} e^{-\delta \tilde{\alpha}_{sw,d}  \rho} \times\\
  e^{-j(k_d\rho \sin\theta_0 \cos (\phi - \phi_0) + l\phi)}
    \end{split}
  \label{eq:Eap_y}
  \end{equation}
  where $l$ is an integer and determines the topological charge (or vorticity state) of the radiated wave. Note that $k_d$ and $\beta_{sw,d}$ are evaluated at the design frequency $f = f_d$, which can be obtained as \cite{minatti2016_FO}:
  \begin{equation}
  k_d = 2\pi f_d \sqrt{\mu_0 \epsilon_0}
  \end{equation}
  \begin{equation}
  \beta_{sw,d} = k_d\sqrt{1 + \frac{\epsilon_0}{\mu_0}X_{op}^2}
  \end{equation}
  where $X_{op}$ is the opaque average reactance and can be calculated by solving the following nonlinear equation:
  \begin{equation}
  \bar{X_\rho} = X_{op}[1 - \frac{X_{op}\epsilon_0\epsilon_r \cot(k_dh\sqrt{\epsilon_r - 1 - \epsilon_0 X_{op}^2/\mu_0})}{\mu_0\sqrt{\epsilon_r - 1 - \epsilon_0 X_{op}^2/\mu_0}}]^{-1}
  \end{equation}
  The leakage factor $ \delta \alpha_{sw,d}$ and propagation constant deviation $\delta \beta_{sw, d}$ can also be calculated by the Floquet-wave expansion method, according to the computation procedure in \cite{minatti2016_FO}. 
Fig. \ref{fig:color_map_alpha_beta} shows the color-map of the attenuation and phase constants in the presence of apace modulation. Observe that the phase constant is almost independent of the modulation index ($M_x$). Fig. \ref{fig:error} presents the calculation error which indicates the sufficient accuracy of the solutions.

\begin{figure}
    \centering
    \includegraphics[width = 0.5\textwidth]{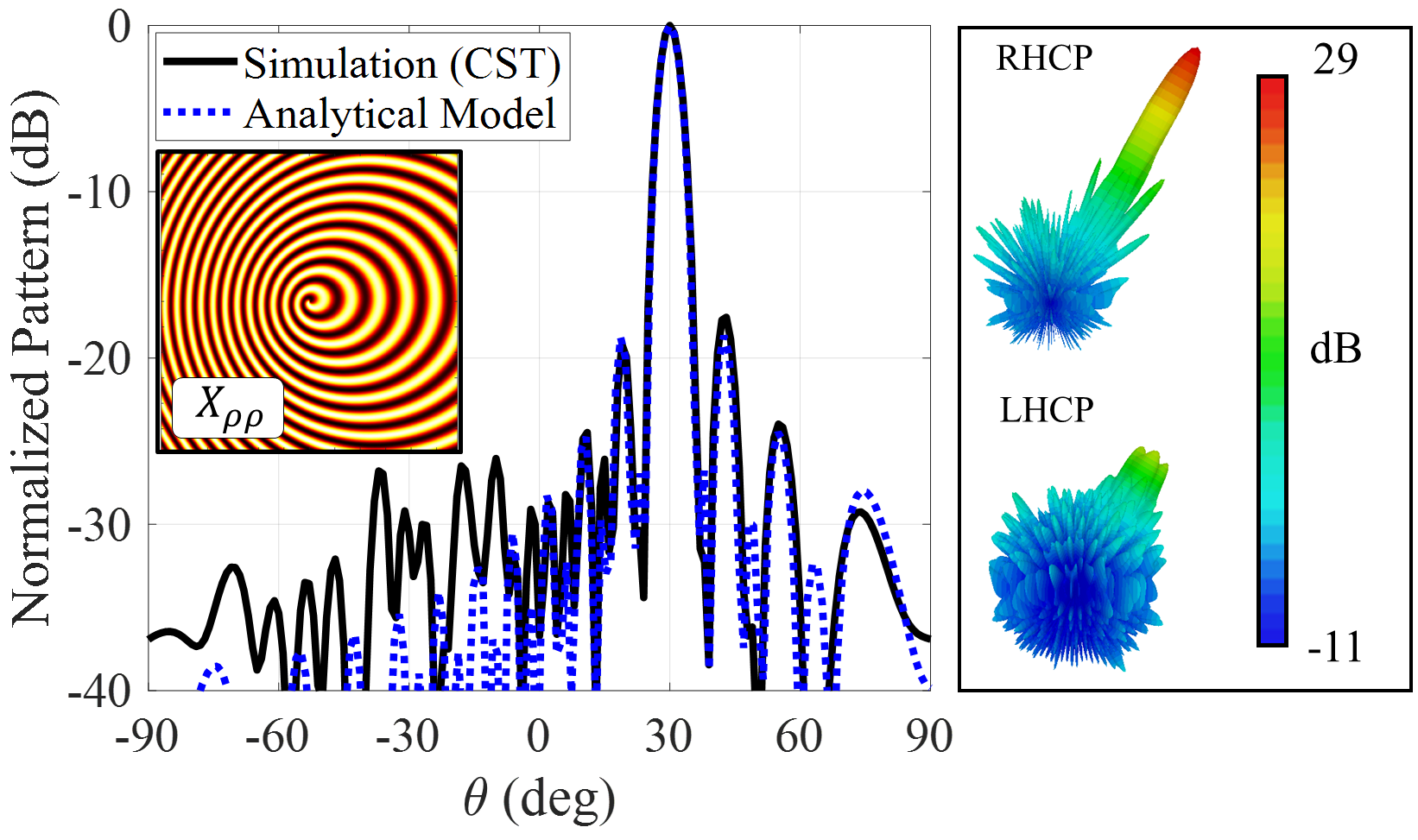}
    \caption{Simulation and analytical results of the circularly polarized pure-space modulated hologram at 18 GHz. Rogers RO4003 ($\epsilon_r = 3.55$, $\tan \delta = 0.0027$) with thickness of $h$ = 1.524 mm is selected as dielectric host medium.}
    \label{fig:Pattern_Static}
  \end{figure}

The ratio between the x and y components of the aperture field indicates the spin angular momentum (polarization) state of the radiated beam. Substituting (\ref{eq:Eap_x}) and (\ref{eq:Eap_y}) in (\ref{eq:Zs_Eap}) and using the asymptotic form of the Hankel function yields:
  \begin{equation}
  \bar{Ks}(\vec{\rho}) = (\beta_{sw,d} + \delta \tilde{\beta}_{sw,d})\rho - k_d\sin \theta_0 \cos (\phi - \phi_0) \rho
  \end{equation}
  \begin{equation}
  \Phi_\rho(\vec{\rho}) = -l\phi + arg\{M_x(\vec{\rho}) \cos \phi + M_y(\vec{\rho}) \sin \phi\}
  \end{equation}
  \begin{equation}
  \Phi_\phi(\vec{\rho}) = -l\phi + arg\{-M_x(\vec{\rho}) \sin \phi + M_y(\vec{\rho}) \cos \phi\}
  \end{equation}
  \begin{equation}
  m_\rho(\vec{\rho}) = |M_x(\vec{\rho}) \cos \phi + M_y(\vec{\rho}) \sin \phi|
  \end{equation}
  \begin{equation}
  m_\phi(\vec{\rho}) = |-M_x(\vec{\rho}) \sin \phi + M_y(\vec{\rho}) \cos \phi|
  \end{equation}
  To evaluate the accuracy of the Floquet-wave expansion, a pure-space modulated surface generating right-hand circular polarization wave is designed. 
  The excitation frequency is 18 GHz and the radiated wave is specified as pencil-shaped directed along $\theta_0 = 30^\circ$ zenith angle. 
  Asymmetric rectangular patches printed on a grounded dielectric \cite{amini2021} are used as constituent pixels to realize the impedance distribution. To cover the required impedance range the unit cell period is selected 2.8 mm ($\approx \lambda / 6$).
  The simulation results for far-field patterns are shown in Fig. \ref{fig:Pattern_Static}. The numerical analysis is performed by the transient solver in CST software \cite{CST}. Observe that the adiabatic Floquet-wave expansion method is sufficiently accurate to analyze the structure. Therefore, relying on the accuracy of this method, we can also analyze spatio-temporally modulated structures.
  \subsection{Spatio-temporally modulated hologram \label{subsection:TV_hologram}}
\begin{figure}
    \centering
         \subfloat[\label{fig:Fig3a}]{%
           \includegraphics[width=0.4\textwidth]{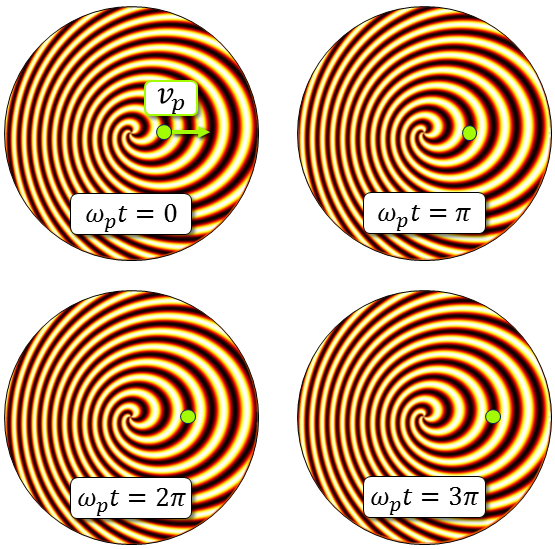}}\\
        \subfloat[\label{fig:Fig3b}]{%
           \includegraphics[width=0.35\textwidth]{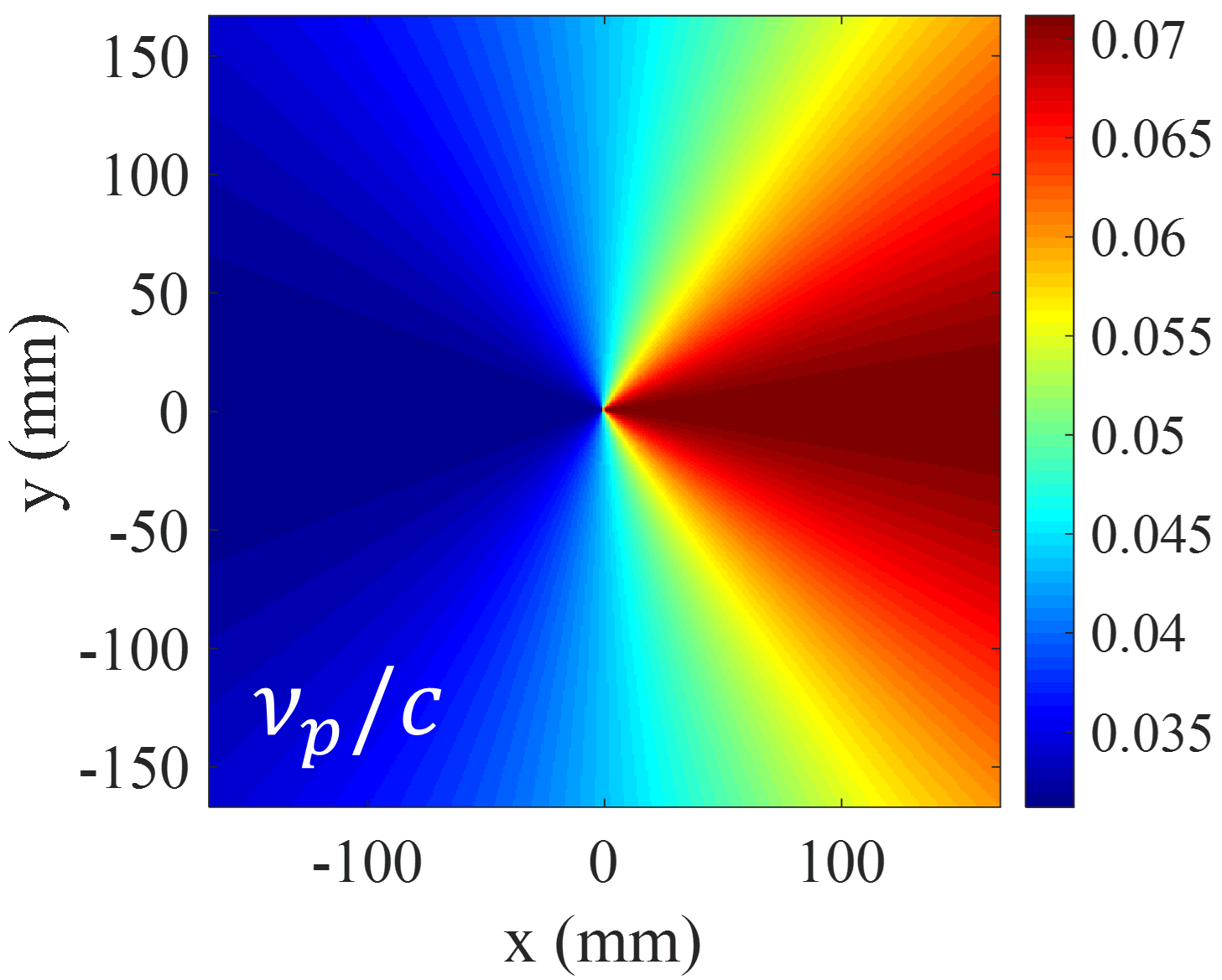}}
      \caption{(a) Impedance distribution for different time slots. The circular indicator shows the movement of the surface impedance. (b) Magnitude of modulation velocity ($\nu_p$).}
  \end{figure}

  Time dependency may be applied to the modulation phase in various ways. However, for simplicity, we assume that the temporal variation of the modulation phase is linear. 
  Therefore, the rapidly varying term $Ks(\vec{\rho}, t)$ may be defined as follows:
\begin{equation}
Ks(\vec{\rho}, t) = \beta_{sw,d}\rho - k_d\sin \theta_0 \cos (\phi - \phi_0) \rho - \omega_p t
\label{eq:Ks_TV}
\end{equation}
where $\omega_p = 2 \pi f_p$ indicates the modulation angular frequency of the boundary condition. 
The phase velocity of the modulation in the $\hat{\rho}$ direction may also be defined as:
\begin{equation}
  v_p = \frac{ \omega_p}{\nabla_{\vec{\rho}}Ks(\vec{\rho}, t).\hat{\rho}}
  \end{equation}
  Fig. \ref{fig:Fig3a} shows the conceptual schematic diagram of the surface impedance for different time slots. The variation of modulation velocity is also depicted in Fig. \ref{fig:Fig3b}. Observe that the temporal change causes the movement of modulation.
  According to (\ref{eq:k_n}) and (\ref{eq:omega_n}), we may obtain the spatial and temporal frequencies for the $n$'th Floquet mode:
  \begin{equation}
  \begin{split}
  \vec{k}^{(n)} = \nabla_{\vec{\rho}}(\tilde{k}_{sw}\rho) + n\beta_{sw,d}\hat{\rho} -
  n k_d \sin \theta_0 \cos (\phi - \phi_0)\hat{\rho} +\\
   n k_d \sin \theta_0 \sin (\phi - \phi_0) \hat{\phi}
  \end{split}
  \label{eq:k_n_tv}
  \end{equation}
  \begin{equation}
  \omega^{(n)} = \omega + n \omega_p 
  \end{equation}
  In (\ref{eq:k_n_tv}), the complex wave number ($\tilde{k}_{sw}$) is unknown, and should be precisely evaluated. A rigorous method for determining the propagation characteristics of such structures is the Eigenmode method, which will be described in the following.

  \section{Calculation of propagation characteristics}
  \subsection{Generalized adiabatic Floquet-wave expansion and Eigenmode equations}
 To establish the Eigenmode equations, we rewrite the transparent boundary condition which was described by (\ref{eq:imp_trans}):
  \begin{equation}
    \vec{E}_t(\vec{\rho}, t) = \sum_n \underline{\underline{Z}}_s(\vec{\rho}, t) \cdot \vec{J}^{(n)}(\vec{\rho}, t)
    \label{eq:Et_expanded}
  \end{equation}
  Using (\ref{eq:Z_s})-(\ref{eq:X_phi_phi}) and applying some mathematical simplifications, we may rewrite (\ref{eq:Et_expanded}) in the form of:
  \begin{equation}
    \begin{split}
      \vec{E}_t(\vec{\rho}, t) = \sum_n \vec{E}^{(n)}(\vec{\rho}, t) = \\
      j \sum_n(\underline{\underline{X}}^{(0)} + \underline{\underline{X}}^{(-1)} + \underline{\underline{X}}^{(+1)}) \cdot \vec{J}^{(n)}(\vec{\rho}, t)
    \end{split}
    \label{eq:Et_En}
  \end{equation}
  where
  \begin{equation}
    \underline{\underline{X}}^{(0)} = \bar{X}_\rho \hat{\rho}\hat{\rho} + \bar{X}_\phi \hat{\phi}\hat{\phi}
  \end{equation}
  \begin{equation}
    \begin{split}
    \underline{\underline{X}}^{(\mp 1)} = \frac{1}{2}m_\rho(\vec{\rho})(\bar{X}_\rho \hat{\rho}\hat{\rho} - \bar{X}_\phi \hat{\phi}\hat{\phi}) e^{\pm j Ks(\vec{\rho}, t)} e^{\pm j \Phi_\rho (\vec{\rho})} + \\ 
    \frac{1}{2}m_\phi(\vec{\rho})(\bar{X}_\rho \hat{\rho}\hat{\phi} + \bar{X}_\rho \hat{\phi}\hat{\rho}) e^{\pm j Ks(\vec{\rho}, t)} e^{\pm j \Phi_\phi (\vec{\rho})}
    \end{split}
    \label{eq:X_mp}
  \end{equation}
  According to (\ref{eq:Et_En}) and (\ref{eq:X_mp}) we may infer that the $n$-indexed electric Floquet mode ($\vec{E}^{(n)}(\vec{\rho}, t)$) should have phase variation as $e^{-jnKs(\vec{\rho}, t)}$. Thus, the components of (\ref{eq:Et_En}) should be rearranged in order to construct the mentioned phase for the electric field.
  \begin{equation}
    \begin{split}
      \vec{E}^{(n)}(\vec{\rho}, t) = j[\underline{\underline{X}}^{(0)} \cdot \vec{J}^{(n)}(\vec{\rho}, t) + \underline{\underline{X}}^{(-1)} \cdot \vec{J}^{(n+1)}(\vec{\rho}, t) + \\
      \underline{\underline{X}}^{(+1)} \cdot \vec{J}^{(n-1)}(\vec{\rho}, t)]
    \end{split}
    \label{eq:En_Jn}
  \end{equation}
  Meanwhile, to complete the equation, the electric field modes must be written in terms of the generalized Green's function of the temporally modulated grounded slab, that is
  \begin{equation}
    \vec{E}^{(n)}(\vec{\rho}, t) \approx \underline{\underline{Z}}_{GF}(\vec{k}^{(n)}, \omega^{(n)}) \cdot \vec{J}^{(n)}(\vec{\rho}, t)
    \label{eq:En_ZGF}
  \end{equation}
  Note that $\vec{k}^{(n)}$ and $\omega^{(n)}$ are spatial and temporal frequencies in the spectral domain, respectively. The calculation procedure of the generalized Green's function is given in Appendix A. Substituting (\ref{eq:En_ZGF}) in (\ref{eq:En_Jn}) gives the Eigenmode equation set:
 \begin{equation}
  \begin{split}
   (-\underline{\underline{Z}}_{GF}(\vec{k}^{(n)}, \omega^{(n)}) + j\underline{\underline{X}}^{(0)}) \cdot \vec{J}^{(n)}(\vec{\rho}, t) + \\
   j\underline{\underline{X}}^{(-1)} \cdot \vec{J}^{(n+1)}(\vec{\rho}, t) + 
   j\underline{\underline{X}}^{(+1)} \cdot \vec{J}^{(n-1)}(\vec{\rho}, t) = 0
  \end{split}
  \label{eq:dispersion_equation}
 \end{equation}
 The above recursive equation relates the (n-1), n and (n+1)-indexed Floquet modes. In order to solve these Eigenmode equations, we must safely truncate the propagation Floquet modes.
In \cite{minatti2016_FO} and \cite{casaletti2019} it has been shown that for leaky-wave structures we can consider only three modes (namely $n = 0, \pm 1$) to obtain sufficiently accurate solutions. By applying some mathematical simplifications and considering mode numbers of $n = 0, \pm 1$, we may rewrite the equation in terms of 0-indexed Floquet mode as follows:
\begin{equation}
  \begin{split}
  (\underline{\underline{X}}_s(\vec{k}^{(0)}, \omega^{(0)}) - \underline{\underline{X}}^{(-1)} \cdot [\underline{\underline{X}}_s(\vec{k}^{(+1)}, \omega^{(+1)})]^{-1} \cdot \underline{\underline{X}}^{(+1)}     
  - \\
  \underline{\underline{X}}^{(+1)} \cdot [\underline{\underline{X}}_s(\vec{k}^{(-1)}, \omega^{(-1)})]^{-1} \cdot \underline{\underline{X}}^{(-1)}) \cdot \vec{J}^{(0)} = 0
  \end{split}
  \label{eq:Xs_eigenmode}
\end{equation}
where
\begin{equation}
  \underline{\underline{X}}_s(\vec{k}^{(n)}, \omega^{(n)}) = j\underline{\underline{Z}}_{GF}(\vec{k}^{(n)}, \omega^{(n)}) + \underline{\underline{X}}^{(0)}
\end{equation}
Considering equation (\ref{eq:Xs_eigenmode}), the nontrivial solutions (eigenvalues) can  be obtained by:
\begin{equation}
  \begin{split}
  det\{(\underline{\underline{X}}_s(\vec{k}^{(0)}, \omega^{(0)}) - \underline{\underline{X}}^{(-1)} \cdot [\underline{\underline{X}}_s(\vec{k}^{(+1)}, \omega^{(+1)})]^{-1} \cdot \underline{\underline{X}}^{(+1)}     
  - \\
  \underline{\underline{X}}^{(+1)} \cdot [\underline{\underline{X}}_s(\vec{k}^{(-1)}, \omega^{(-1)})]^{-1} \cdot \underline{\underline{X}}^{(-1)})\}  = 0
  \end{split}
  \label{eq:det}
\end{equation}
Finally, the leaky-mode ((-1)-indexed mode) field can also be calculated as follows:
\begin{equation}
  \vec{E}^{(-1)} = j[\underline{\underline{X}}^{(-1)} - \underline{\underline{X}}^{(0)}\cdot (\underline{\underline{X}}_s(\vec{k}^{(-1)}, \omega^{(-1)}))^{-1}] \cdot \vec{J}^{(0)}
\end{equation}
\begin{figure*}
  \centering
       \subfloat[\label{fig:omega_p1}]{%
         \includegraphics[width=0.33\textwidth]{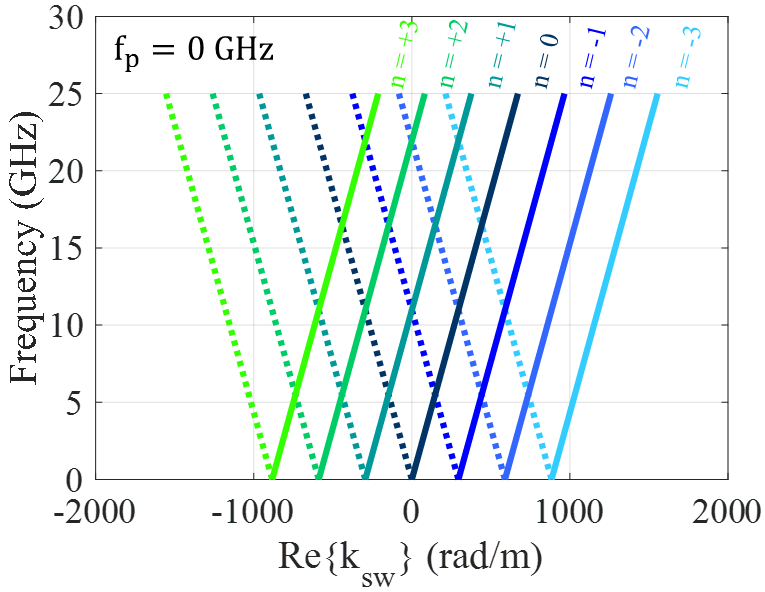}}
      \subfloat[\label{fig:omega_p2}]{%
         \includegraphics[width=0.33\textwidth]{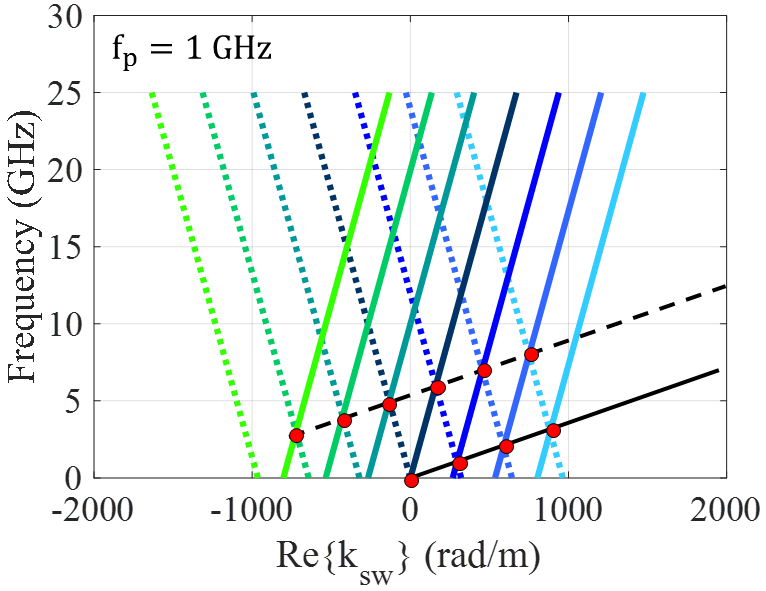}}
      \subfloat[\label{fig:omega_p}]{%
         \includegraphics[width=0.33\textwidth]{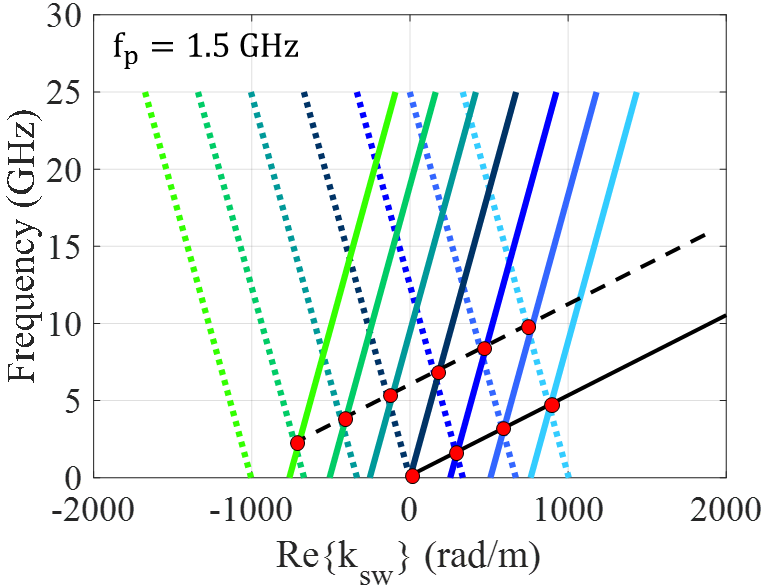}}
    \caption{Asymptotic dispersion curves in $\phi = 0^\circ$ for different values of $f_p$. (a) $f_p$ = 0 GHz (pure-space modulation). (b) $f_p$ = 1 GHz. (c) $f_p$ = 1.5 GHz.}
    \label{fig:dispersion_asymptotic}
\end{figure*}
\subsection{Calculating eigenstates for the asymptotic case ($m_\rho, m_\phi \rightarrow 0$)}
In order to accurately solve the nonlinear equation in (\ref{eq:det}), we need to estimate an initial guess. The initial guesses can be obtained by assuming an asymptotic case such that the modulation indexes tend to zero ($m_\rho, m_\phi \rightarrow 0$). In this situation, the effect of coefficients $\underline{\underline{X}}^{(-1)}$ and $\underline{\underline{X}}^{(+1)}$ are negligible, and they can be considered equal to zero. Therefore, (\ref{eq:dispersion_equation}) can be deduced as follows:
\begin{equation}
  (-\underline{\underline{Z}}_{GF}(\vec{k}^{(n)}_{asm}, \omega^{(n)}) + j\underline{\underline{X}}^{(0)}) \cdot \vec{J}^{(n)}(\vec{\rho}, t) = 0
\end{equation}
Note that, in the above equation, the subscript $"asm"$ indicates the wavenumber in the asymptotic form.
In order to have nontrivial solutions, the following relation must be satisfied:
\begin{equation}
  det\{(-\underline{\underline{Z}}_{GF}(\vec{k}^{(n)}_{asm}, \omega^{(n)}) + j\underline{\underline{X}}^{(0)})\} = 0
\end{equation}
Given that the modulation indexes tend to zero, we can consider the leakage constant (the imaginary part of n-indexed wave-vector) to be negligible. Using some mathematical simplifications, we obtain:
\begin{equation}
  \beta^{(n)}_{asm}\hat{\rho} \approx \frac{\omega + n\omega_p}{c} \sqrt{1 + (\frac{X_{op}}{\eta_0})^2}\hat{\rho}
  \label{eq:beta_n_asymptotic}
\end{equation}
Substituting (\ref{eq:beta_n_asymptotic}) in (\ref{eq:k_n}) yields
\begin{equation}
  \nabla_{\vec{\rho}}(\tilde{k}_{sw, asm}\rho) = - \nabla_{\vec{\rho}}[nKs(\vec{\rho}, t)] \pm \frac{\omega + n\omega_p}{c}\sqrt{1 + (\frac{X_{op}}{\eta_0})^2} \hat{\rho}
  \label{eq:dispersion_TV}
\end{equation}
where $+$ and $-$ signs indicate the forward and backward travelling wave solutions for each value of $n$.
The above equation expresses the dispersion characteristics of a surface wave guiding structure for the limit of $m_\rho, m_\phi \rightarrow 0$. In the general case, $\tilde{k}_{sw}$ is a function of $\rho$ and $\phi$. Therefore, equation (\ref{eq:dispersion_TV}) may be expanded as follows:
\begin{equation}
  \begin{split}
  \frac{\partial}{\partial \rho} (\tilde{k}_{sw, asm} \rho) = -n(\beta_{sw, d} - k_d \sin \theta_0 \cos (\phi - \phi_0)) \\
  \pm \frac{\omega + n\omega_p}{c}\sqrt{1 + (\frac{X_{op}}{\eta_0})^2}
  \end{split}
  \label{eq:ksw_rho}
\end{equation}
\begin{equation}
  \frac{\partial}{\partial \phi} (\tilde{k}_{sw, asm}) = -nk_d \sin \theta_0 \sin (\phi - \phi_0)
  \label{eq:ksw_phi}
\end{equation}
Note that, equations (\ref{eq:ksw_rho}) and (\ref{eq:ksw_phi}) are used to determine the exact values of higher order eigenstates.

To observe the behavior of the dispersion curves in the presence of space-time modulation, they are plotted in 
Fig. \ref{fig:dispersion_asymptotic} for different values of $f_p$ (The observation plane is $\phi = 0^\circ$).
The antenna is specified to radiate in the direction $\theta_0 = 30^\circ, \phi_0 = 0^\circ$  at 18 GHz. The opaque average reactance $X_{op}$ is also chosen as $0.76\eta_0$.
The solid curves represent forward-travelling solutions and the dashed ones are related to the backward-travelling modes. 
Observe that, by applying temporal modulation, an asymmetry with respect to the vertical axis appears in the curves justifying the nonreciprocal response of temporally modulated structures. Note that by increasing $\omega_p$ the curves are displaced in an asymmetrical manner, and the degree of nonreciprocity increases. It can also be seen that in the presence of modulation, the Brillouin zone becomes parallel to the line with slope of $\nu_p/c$. 
\begin{figure}
  \centering
  \includegraphics[width = 0.32\textwidth]{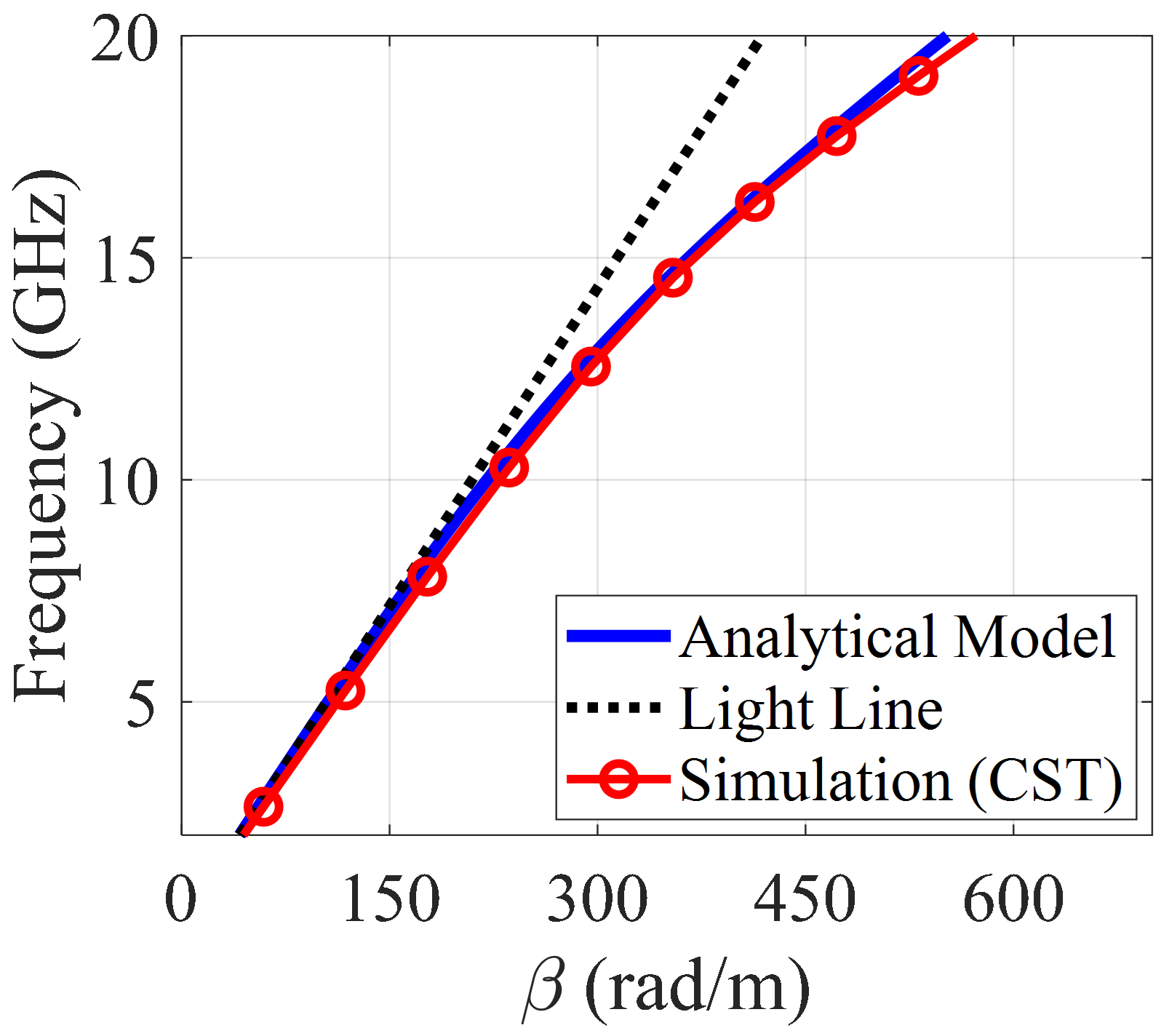}
  \caption{Comparison between numerical and analytical results of the dispersion curve for an unmodulated surface waveguide with average transparent reactance of $X_\rho = -0.9\eta_0$.}
  \label{fig:dispersion_sim_analytical}
\end{figure}
\begin{figure*}
  \centering
      \subfloat[\label{fig:dispersion_Tx}]
      {%
         \includegraphics[width=0.4\textwidth]{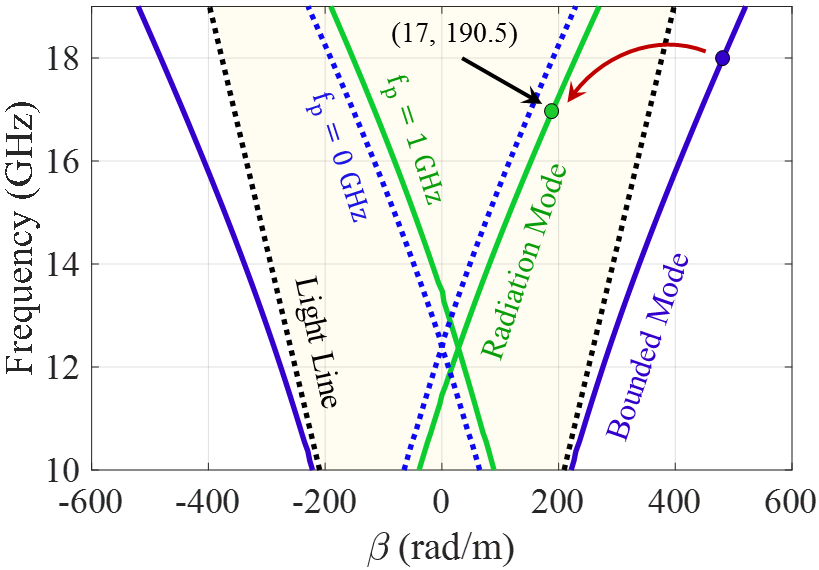}
      }
      \subfloat[\label{fig:dispersion_Rx}]
      {%
         \includegraphics[width=0.4\textwidth]{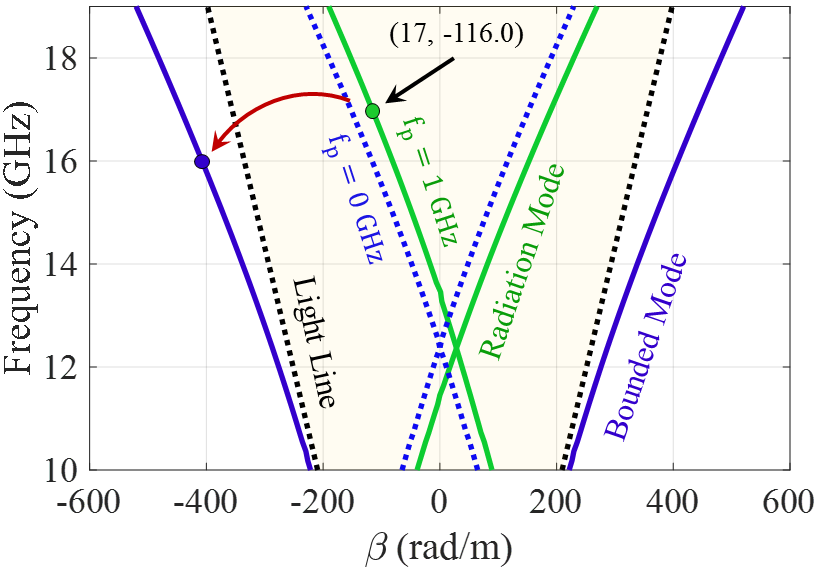}
      }
    \caption{Dispersion curves of bounded and radiative harmonics in the (a) transmission and (b) reception modes. Dotted lines represent the dispersion curves for the pure-space modulated hologram ($f_p = 0$). Asymmetric displacement of the dispersion curves can be observed for $f_p  \neq 0$.}
    \label{fig:dispersion_Tx_Rx}
\end{figure*}
\subsection{Calculating eigenstates in the presence of space and time modulations}
The dispersion curves obtained in Fig \ref{fig:dispersion_asymptotic} can be used as initial guesses to determine the accurate solutions of  (\ref{eq:det}). 
To investigate the accuracy of the proposed analytical method, the dispersion diagram calculated for a surface waveguide is compared with the corresponding numerical results in Fig. \ref{fig:dispersion_sim_analytical}.
The modulation indexes in the $x$ and $y$ directions as well as the average reactance are supposed to be $M_x = 0$, $M_y = 0$ and $X_\rho = -0.9\eta_0$, respectively.
Observe that the numerical and analytical results are in good agreement.

\begin{figure}
  \centering
  \includegraphics[width = 0.42\textwidth]{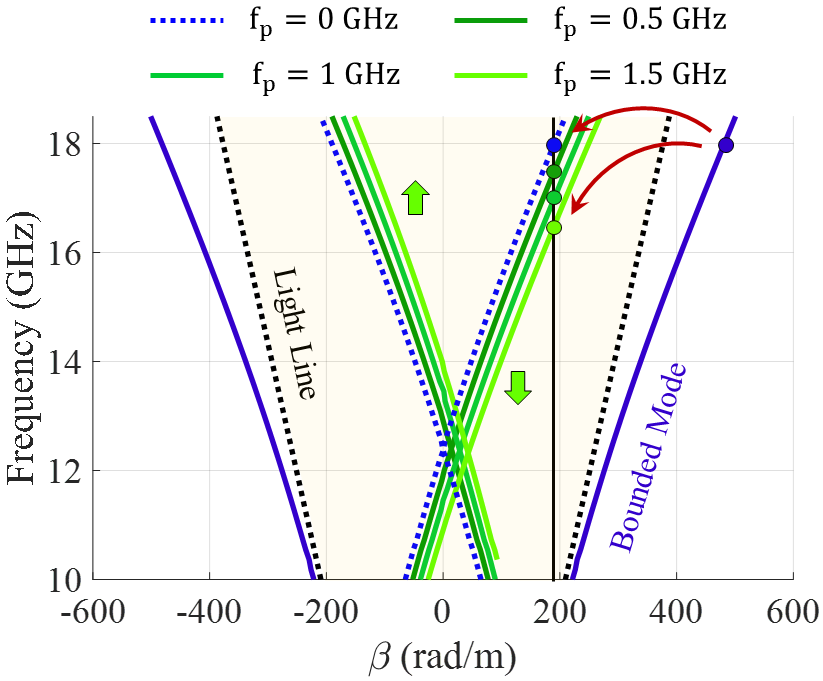}
  \caption{Dispersion diagram of the proposed hologram for different values of $f_p$.}
  \label{fig:dispersion_2D_scan}
\end{figure}

Imposition of the space-time modulation results in the following:
\begin{enumerate}
  \item The higher-order Floquet modes contribute to the propagation. In this case, the energy carried by the surface wave may be coupled to the radiation mode with the index number $n = -1$.
  \item Due to the temporal modulation, the dispersion curves corresponding to the positive (negative) harmonics move upwards (downwards). The amount of these displacements depends on the pumping temporal frequency ($f_p$) \cite{correas2015}.
\end{enumerate}
Fig. \ref{fig:dispersion_Tx_Rx} shows the dispersion solutions for bounded and radiative harmonics in the case of $X_\rho = -0.9\eta_0$ and $M_x = 0.3$. The pumping frequency $f_p$ is specified at 1 GHz. The parameters $\theta_0$ and $\phi_0$ in (\ref{eq:Ks_TV}) are selected as $30^\circ$ and $0^\circ$, respectively. The dotted blue lines represent the dispersion curves for the pure-space modulation case. Observe that, injecting temporal modulation imposes an asymmetric shifting on the curves, leading to nonreciprocal response and Doppler-shift effect. 
To better understand its nonreciprocal behavior, we assume that the antenna is in transmission mode and the excitation frequency is 18 GHz (see Fig. \ref{fig:dispersion_Tx}).
According to (\ref{eq:omega_n}), the frequency of surface mode (0-indexed mode) is equal to the source frequency (18 GHz). For the pure-space modulation, the (-1)-indexed phase constant at this frequency is  $\beta^{(-1)}\approx 190.5$ (rad/m). However, in the presence of time modulation, the frequency corresponding to this phase constant is no longer equal to 18 GHz (due to the curve displacement). Results in Fig. \ref{fig:dispersion_Tx} indicate that the (-1)-indexed mode radiates at 17 GHz.
Therefore, its corresponding beam angle obtained as:
\begin{equation}
  \theta_{Tx} = \sin ^{-1}(\frac{\beta^{(-1)}}{k^{(-1)}}) = \sin ^{-1}(\frac{c \beta^{(-1)}}{2\pi(f - f_p)}) \approx 32.3^\circ
\end{equation} 
In the reception case, the incoming wave is plane wave at 17 GHz. In this case the maximum coupling occurs at $\beta = -116$ (rad/m) with incoming angle of $\theta_{Rx} \approx 19^\circ$ and the 0-indexed mode is excited at 16 GHz, as shown in Fig. \ref{fig:dispersion_Rx}. Therefore, no power is received at 18 GHz, showing a nonreciprocal behavior by this hologram.
Note that the proposed antenna also has the scanning capability. The mechanism of beam scanning in this structure is different from pure-space holograms, where the beam scanning can be achieved by changing the excitation frequency, whereas in time modulated holograms, the modulation frequency can be used as a degree of freedom to control the beam direction. Fig. \ref{fig:dispersion_2D_scan} shows the dispersion curves for different values of $f_p$. Observe that, as $f_p$ increases, the eigenstate (beam direction) tends toward the light line (end-fire direction). 
\begin{figure*}
  \centering
      \subfloat[\label{fig:RHCP_LHCP_fp_0_5GHz}]
      {%
         \includegraphics[width=0.32\textwidth]{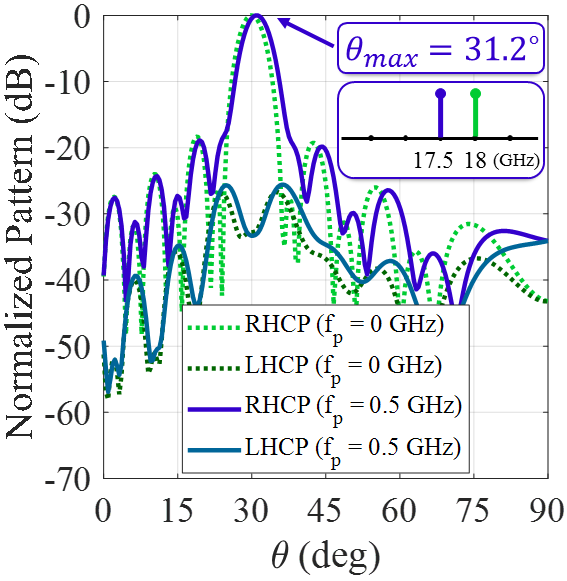}
      }
      \subfloat[\label{fig:RHCP_fp_1GHz}]
      {%
         \includegraphics[width=0.32\textwidth]{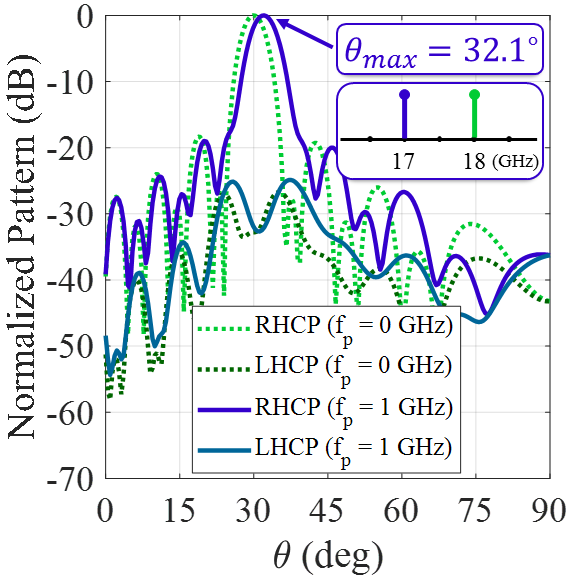}
      }
      \subfloat[\label{fig:RHCP_fp_1_5GHz}]
      {%
         \includegraphics[width=0.32\textwidth]{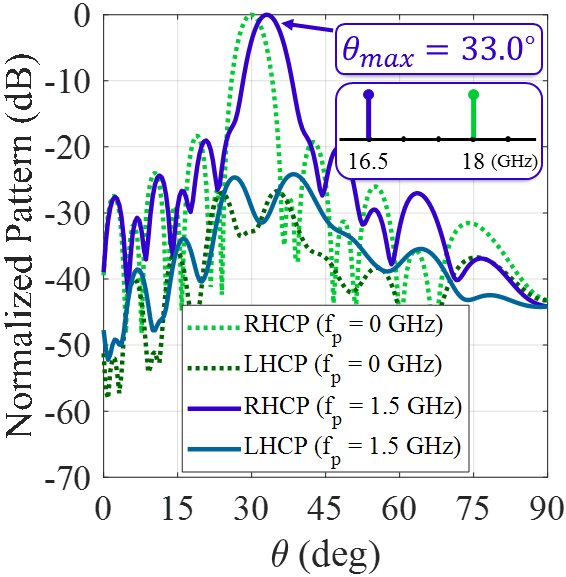}
      }
    \caption{2-D normalized patterns of RHCP and LHPC components for different values of $f_p$. (a) $f_p = 0.5$ GHz. (b) $f_p = 1$ GHz. (c) $f_p = 1.5$ GHz. Note that the radiation frequency is $f_0 - f_p$.}
    \label{fig:RHCP_LHCP}
\end{figure*}
\begin{figure*}
  \centering
      \subfloat[\label{fig:RHCP_uv_fp_0_5GHz}]
      {%
         \includegraphics[width=0.33\textwidth]{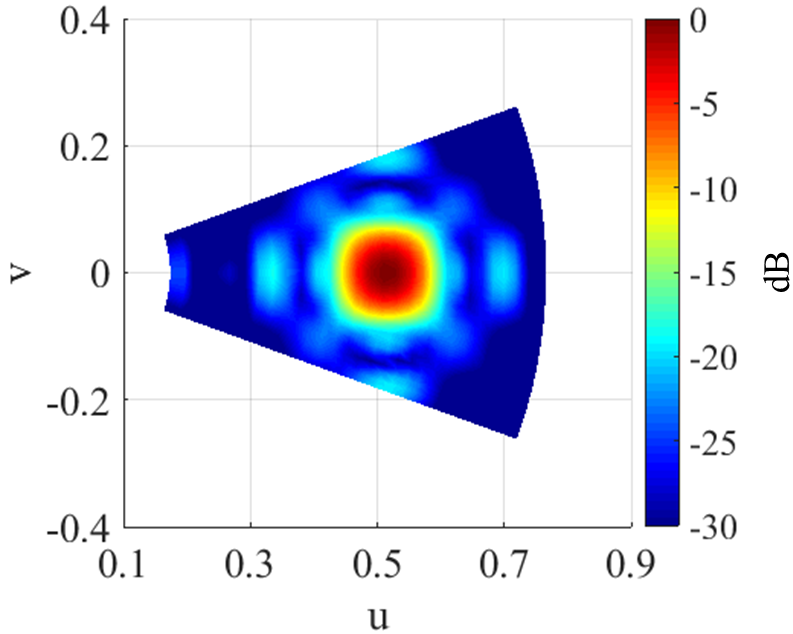}
      }
      \subfloat[\label{fig:RHCP_uv_fp_1GHz}]
      {%
         \includegraphics[width=0.33\textwidth]{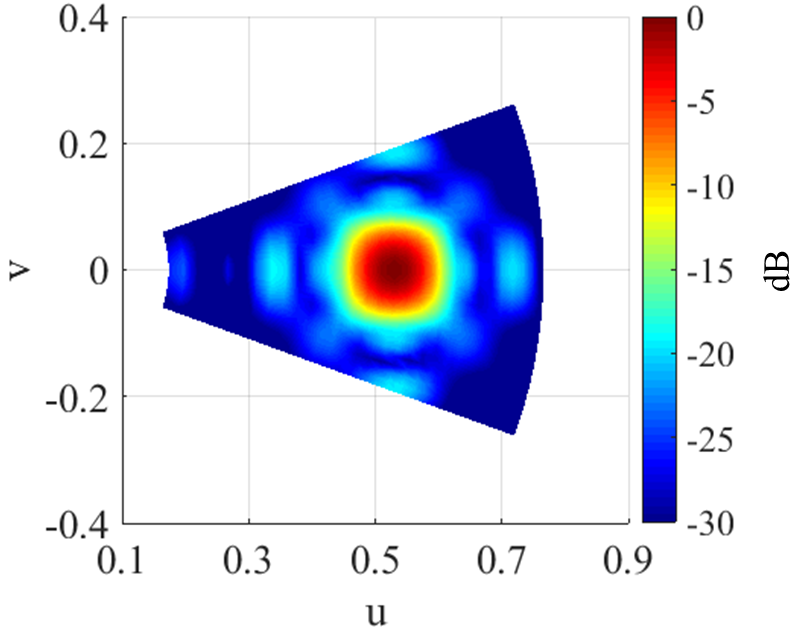}
      }
      \subfloat[\label{fig:RHCP_uv_fp_1_5GHz}]
      {%
         \includegraphics[width=0.33\textwidth]{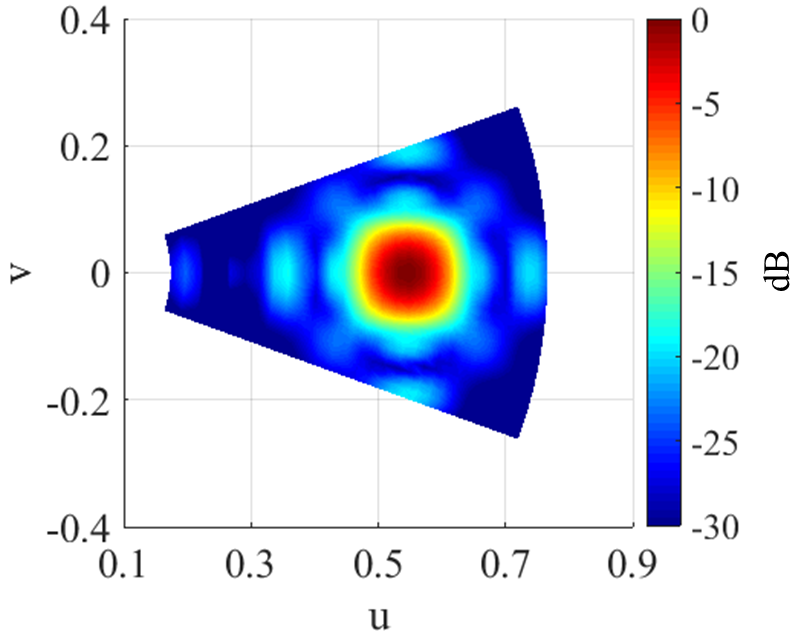}
      }
    \caption{Far-field distributions of RHCP components for different values of $f_p$. (a) $f_p = 0.5$ GHz. (b) $f_p = 1$ GHz. (c) $f_p = 1.5$ GHz.}
    \label{fig:RHCP_uv}
\end{figure*}
\section{Calculating far-field pattern}
To obtain the radiation characteristics from $n$-indexed Floquet mode, the field equivalence principle may be used with the Fourier transformation equations \cite{balanis2016}:
\begin{equation}
  \vec{E}_{FF}^{(n)}(r, \theta, \phi)\approx \frac{jk^{(n)}e^{-jk^{(n)}r}}{2\pi r}[F_\theta^{(n)}(\theta, \phi) \hat{\theta} + F_\phi^{(n)}(\theta, \phi)\hat{\phi}]
\end{equation}
where
\begin{equation}
  F_\theta^{(n)}(\theta, \phi) = \tilde{E}_x^{(n)} \cos \phi + \tilde{E}_y^{(n)} \sin \phi
\end{equation}
\begin{equation}
  F_\phi^{(n)}(\theta, \phi) = \cos \theta (-\tilde{E}_x^{(n)} \sin \phi + \tilde{E}_y^{(n)} \cos \phi)
\end{equation}
Note that, $\tilde{E}_x^{(n)}$ and $\tilde{E}_y^{(n)}$ indicate the far-zone components of electric field which can be expressed as follows:
\begin{equation}
  \tilde{E}_x^{(n)}(k_x, k_y) = \iint_{ap} \vec{E}^{(n)}(\rho', \phi')\cdot \hat{x} e^{j(k_x x' + k_y y'}\rho' d\rho' d\phi'
\end{equation}
\begin{equation}
  \tilde{E}_y^{(n)}(k_x, k_y) = \iint_{ap} \vec{E}^{(n)}(\rho', \phi')\cdot \hat{y} e^{j(k_x x' + k_y y'}\rho' d\rho' d\phi'
\end{equation}
where $x' = \rho' \cos \phi '$ and $y' = \rho' \sin \phi'$.
Note that, $k_x$ and $k_y$ are the x and y components of $n$-indexed wavenumber, respectively, being defined as
\begin{equation}
  k_x = k^{(n)} \sin \theta \cos \phi = \frac{2\pi(f + n f_p)}{c} \sin \theta \cos \phi
\end{equation}
\begin{equation}
  k_y = k^{(n)} \sin \theta \sin \phi = \frac{2\pi(f + n f_p)}{c} \sin \theta \sin \phi
\end{equation}
The radiation frequency is $f + n f_p$ indicating a Doppler-shift effect on the radiative modes.
Fig. \ref{fig:RHCP_LHCP} show the  (-1)-indexed far-field pattern in the $\phi = 0^\circ$ plane for different values of $f_p$. 
The radiated beams for $f_p = 0.5, 1.0,$ and $1.5$ GHz, are directed along $\theta_{max} = 31.2^\circ$, $32.1^\circ$ and $33.0^\circ$, respectively.
 Observe that the co- to cross-polarization has an acceptable level (> 24 dB) with right-handed circular polarization. 
 Fig. \ref{fig:RHCP_uv} shows the far-field distributions in the $u-v$ plane. Results indicate that, in the presence of time modulation, the beam has retained its pencil shape.

\section{Conclusion}
In conclusion, we presented the generalized adiabatic Floquet-wave expansion method for the design of temporally modulated leaky-wave holograms. This method may be exploited for synthesizing arbitrary objective waves with desired spin angular momentum states. Using aperture field estimation method, we can readily provide a systematic approach for designing two-dimensional leaky-wave holographic antennas with nonreciprocal behavior. 
 The results also indicated that the Doppler shift is proportional to the temporal modulation frequency, which can be an additional degree of freedom for beam control of leaky-wave holograms.
This framework extensively expands the application range of modulated leaky-wave holograms.

\appendix
\subsection{Green's function of temporally modulated surface waveguide}
The $n$-indexed electric field mode can be approximated through the Green's function of a grounded dielectric slab with thickness of $h$ and dielectric constant of $\epsilon_r$.
Note that for a spatio-temporally modulated impedance boundary condition the spectral variables are $\vec{k}^{(n)}$ and $\omega^{(n)}$.
Therefore, the Green's function in the spectral space has the following form \cite{minatti2016_FO}:
\begin{equation}
  \underline{\underline{Z}}_{GF}(\vec{k}^{(n)}, \omega^{(n)}) = -j[\underline{\underline{X}}_0^{-1}(\vec{k}^{(n)}, \omega^{(n)}) + \underline{\underline{X}}_g^{-1}(\vec{k}^{(n)}, \omega^{(n)})]^{-1}
\end{equation}
where $\underline{\underline{X}}_0(\vec{k}^{(n)}, \omega^{(n)})$ and $\underline{\underline{X}}_g^{-1}(\vec{k}^{(n)}, \omega^{(n)})$ are obtained using the two-port network model presented in \cite{martini2015}.
\begin{equation}
  \begin{split}
  \underline{\underline{X}}_0(\vec{k}^{(n)}, \omega^{(n)}) = -\eta_0 c \frac{\sqrt{\vec{k}^{(n)}\cdot \vec{k}^{(n)} - (\omega^{(n)})^2/c^2}}{\omega^{(n)}}\hat{\rho}\hat{\rho} +\\
  \frac{\eta_0}{c} \frac{\omega^{(n)}}{\sqrt{\vec{k}^{(n)}\cdot \vec{k}^{(n)} - (\omega^{(n)})^2/c^2}} \hat{\phi}\hat{\phi}
  \end{split}
\end{equation}
\begin{equation}
  \begin{split}
  \underline{\underline{X}}_g(\vec{k}^{(n)}, \omega^{(n)}) = [\eta_0 c \frac{\sqrt{\epsilon_r(\omega^{(n)})^2/c^2 - \vec{k}^{(n)}\cdot \vec{k}^{(n)}}}{\epsilon_r \omega^{(n)}}\hat{\rho}\hat{\rho} + \\
  \frac{\eta_0}{c}\frac{\omega^{(n)}}{\sqrt{\epsilon_r(\omega^{(n)})^2/c^2 - \vec{k}^{(n)}\cdot \vec{k}^{(n)}}} \hat{\phi}\hat{\phi}] \times \\
  \tan (h \sqrt{\epsilon_r(\omega^{(n)})^2/c^2 - \vec{k}^{(n)}\cdot \vec{k}^{(n)}})
  \end{split}
\end{equation}
The above equation is similar to the equations obtained for the pure-space modulated surfaces, except that for the space-time modulation, $\omega^{(n)}$ should be used instead of $\omega$.

\ifCLASSOPTIONcaptionsoff
  \newpage
\fi

%







\begin{thebibliography}{1}
  \bibitem{yu2011}
  N. Yu, P. Genevet, M. A. Kats, F. Aieta, J.-P. Tetienne, F. Capasso, andZ. Gaburro,  "Light propagation  with  phase  discontinuities: generalized laws of reflection and refraction," \textit{Science}, p. 1210713, 2011.

  \bibitem{kildishev2013}
A.  V.  Kildishev,  A.  Boltasseva,  and  V.  M.  Shalaev,  "Planar  photonics with metasurfaces," \textit{Science,}  vol. 339, no. 6125, p. 1232009, 2013.

\bibitem{yu2014}
N. Yu and F. Capasso, "Flat optics with designer metasurfaces," \textit{Nature Mater.}, vol. 13, no. 2, pp. 139-150, 2014.

\bibitem{karimi2014}
E. Karimi, S. A. Schulz, I. De Leon, H. Qassim, J. Upham, and R. W.Boyd,  "Generating  optical  orbital  angular  momentum  at  visible  wave-lengths using a plasmonic metasurface," \textit{Light: Science \& Applications}, vol. 3, no. 5, p. e167, 2014.

\bibitem{ding2015}
X. Ding, F. Monticone, K. Zhang, L. Zhang, D. Gao, S. N. Burokur, A.de Lustrac, Q. Wu, C.-W. Qiu, and A. Al`u, "Ultrathin pancharatnam-berry metasurface with maximal cross-polarization efficiency," \textit{Adv. Mater.}, vol.27, pp. 1195-1200, 2015.

\bibitem{estakhri2016}
N.   Mohammadi   Estakhri   and   A.   Alu,   "Wave-front   transformation with  gradient  metasurfaces," \textit{Phys.  Rev.  X},  vol.  6,  no.  4, p. 041008, Oct.  2016.

\bibitem{arbabi2015}
A. Arbabi, Y. Horie, M.  Bagheri and A. Faraon, "Dielectric metasur-faces for complete control of phase and polarization with subwavelength spatial  resolution  and  high  transmission," \textit{Nat.  Nanotech.},  vol.  10,  pp.937-943, 2015.

\bibitem{fong2010}
B. H. Fong, J. S. Colburn, J. J. Ottusch, J. L. Visher, and D. F. Sievenpiper, "Scalar and Tensor Holographic Artificial Impedance Surfaces," \textit{IEEE Trans. Antennas Propag.}, vol. 58, no. 10, pp. 3212-3221, Oct. 2010.

\bibitem{patel2011}
A. M. Patel and A. Grbic, "A printed leaky-wave antenna based on a sinusoidally-modulated reactance surface", \textit{IEEE Trans. Antennas Propag.}, vol. 59, no. 6, pp. 2087-2096, Jun. 2011.

\bibitem{minatti2011}
 G. Minatti, F. Caminita, M. Casaletti and S. Maci, "Spiral Leaky-Wave Antennas Based on Modulated Surface Impedance", \textit{IEEE Trans. Antennas Propag.}, vol. 59, no. 12, pp. 4436-4444, Dec. 2011.
 
 \bibitem{sun2012}
S. L. Sun, Q. He, S. Y. Xiao, Q. Xu, X. Li and L. Zhou, "Gradient-index meta-surfaces as a bridge linking propagating waves and surface waves", \textit{Nat. Mater.}, vol. 11, pp. 426-431, 2012.

\bibitem{minatti2015}
G. Minatti, M. Faenzi, E. Martini, F. Caminita, P. De Vita, D. Gonzalez-Ovejero, M. Sabbadini, and S. Maci, "Modulated Metasurface Antennas for Space: Synthesis, Analysis and Realizations," \textit{IEEE Trans. Antennas Propag.}, vol. 63, no. 4, pp. 1288-1300, Apr. 2015.

\bibitem{li2016}
M. Li, S.-Q. Xiao and D. F. Sievenpiper, "Polarization-insensitive holographic surfaces with broadside radiation", \textit{IEEE Trans. Antennas Propag.}, vol. 64, no. 12, pp. 5272-5280, Dec. 2016.

\bibitem{minatti2016}
G. Minatti, F. Caminita, E. Martini, M. Sabbadini and S. Maci, "Synthesis of modulated-metasurface antennas with amplitude phase and polarization control", \textit{IEEE Trans. Antennas Propag.}, vol. 64, no. 9, pp. 3907-3919, Sep. 2016.

\bibitem{movahhedi2019}
M. Movahhedi, M. Karimipour and N. Komjani, "Multibeam bidirectional wideband/wide-scanning-angle holographic leaky-wave antenna", \textit{IEEE Antennas Wireless Propag. Lett.}, vol. 18, no. 7, pp. 1507-1511, Jul. 2019.

\bibitem{bodehou2019}
M. Bodehou, C. Craeye, E. Martini and I. Huynen, "A quasi-direct method for the surface impedance design of modulated metasurface antennas", \textit{IEEE Trans. Antennas Propag.}, vol. 67, no. 1, pp. 24-36, Jan. 2019.

\bibitem{shaltout2019}
A. M. Shaltout, V. M. Shalaev and M. L. Brongersma, "Spatiotemporal light control with active metasurfaces", \textit{Science}, vol. 364, no. 6441, 2019.

\bibitem{kang2019}
L. Kang, R. P. Jenkins and D. H. Werner, "Recent progress in active optical metasurfaces", \textit{Adv. Opt. Mater.}, vol. 7, no. 14, Jul. 2019.

\bibitem{galiffi2020}
E. Galiffi, Y. -T. Wang, Z. Lim, J. B. Pendry, A Alù, and P. A. Huidobro, "Wood anomalies and surface-wave excitation with a time grating," \textit{Phys. Rev. Lett.}, vol. 125, no. 12, p. 127403, Sep. 2020.

\bibitem{cardin2020}
A. E. Cardin, S. R. Silva, S. R. Vardeny, W. J. Padilla, A. Saxena, A. J. Taylor, W. J. M. Kort-Kamp, H.-T. Chen, D. A. R. Dalvit, and A. K. Azad, "Surface-wave-assisted nonreciprocity in spatio-temporally modulated metasurfaces," \textit{Nat. Commun.}, vol. 11, no. 1, p. 1469, Mar. 2020.

\bibitem{correas2015}
D. Correas-Serrano, J. S, Gomez-Diaz, D. L. Sounas, Y. Hadad, A. Alvarez-Melcon, and A. Alù, "Nonreciprocal graphene devices and antennas based on spatiotemporal modulation",  \textit{IEEE Antennas Wirel. Propag. Lett.}, vol. 15, pp.1529-1532, 2015.

\bibitem{taravati2017_nongyro}
S. Taravati, B. A. Khan, S. Gupta, K.Achouri and C. Caloz, "Nonreciprocal nongyrotropic magnetless metasurface", \textit{IEEE Trans. Antennas Propag.}, vol. 65, no. 7, pp. 3589-3597, Jul. 2017.

\bibitem{ramaccia2019}
    D. Ramaccia, D. L. Sounas, A. Alu, A. Toscano, and F. Bilotti, "Phase-induced frequency conversion and Doppler effect with time-modulated metasurfaces," \textit{IEEE Trans. Antennas Propag.}, vol. 68, no. 3, pp. 1607-1617, 2020.
  
    
    \bibitem{hu2021}
    Y. Hu, M. Tong, Z. Xu, X. Cheng, and T. Jiang, " Spatiotemporal terahertz metasurfaces for ultrafast all-optical switching with electric-triggered bistability", \textit{Laser Photon. Rev.}, vol. 15, no. 3, pp. 2000456, 2021.


    \bibitem{taravati2017_TAP}
    S. Taravati and C. Caloz "Mixer-duplexer-antenna leaky-wave system based on periodic space-time modulation," \textit{IEEE Trans. Antennas Propag.}, vol. 65, no. 2, pp. 442-452, Feb. 2017.

    \bibitem{taravati2020}
S. Taravati and G. V. Eleftheriades, "Space-time medium functions as a perfect antenna-mixer-amplifier transceiver", \textit{Phys. Rev. Appl.}, vol. 14, no. 5, pp. 054017, 2020.

\bibitem{amini2022}
A. Amini, and H. Oraizi, "Temporally modulated one-dimensional leaky-wave holograms," \textit{Sci. Rep.}, vol. 12, no. 1, p. 8488, May 2022.

\bibitem{nayeri2014}
P. Nayeri, M. Liang, R. A. Sabory-Garcia, M. Tuo, F. Yang, M. Gehm, H, Xin, and A. Z. Elsherbeni, "3D printed dielectric reflectarrays: Low-cost high-gain antennas at sub-millimeter waves", \textit{IEEE Trans. Antennas Propag.}, vol. 62, no. 4, pp. 2000-2008, Apr. 2014.

\bibitem{gabor1948}
D. Gabor, "A new microscopic principle", \textit{Nature}, vol. 161, no. 4098, pp. 777-778, 1948.

\bibitem{hariharan1996}
P. Hariharan, \textit{Optical Holography: Principles Techniques and Applications}, U.K., Cambridge:Cambridge Univ. Press, 1996.

\bibitem{teniou2017}
M. Teniou, H. Roussel, N. Capet, G. Piau, and M. Casaletti, "Implementation of radiating aperture field distribution using tensorial metasurfaces," \textit{IEEE Trans. Antennas Propag.},  vol. 65, no. 11, pp. 5895-5907, Nov. 2017.

\bibitem{ovejero2015}
D. González-Ovejero and S. Maci, "Gaussian ring basis functions for the analysis of modulated metasurface antennas", \textit{IEEE Trans. Antennas Propag.}, vol. 63, pp. 3982-3993, Sep. 2015.

\bibitem{bodehou2019_mom}
M. Bodehou, D. González-Ovejero, C. Craeye and I. Huynen, "Method of Moments simulation of modulated metasurface antennas with a set of orthogonal entire-domain basis functions", \textit{IEEE Trans. Antennas Propag.}, vol. 67, no. 2, pp. 1119-1130, 2019.

\bibitem{minatti2016_FO}
G. Minatti, F. Caminita, E. Martini and S. Maci, "Flat optics for leaky-waves on modulated metasurfaces: Adiabatic Floquet-wave analysis", \textit{IEEE Trans. Antennas Propag.}, vol. 64, no. 9, pp. 3896-3906, Sep. 2016.

\bibitem{amini2021}
A. Amini, and H. Oraizi, "Adiabatic Floquet-wave Expansion for the Analysis of Leaky-Wave Holograms Generating Polarized Vortex Beams," \textit{Phys. Rev. Appl.}, vol. 15, p. 034049, 2021.

    
\bibitem{cassedy1965}
E. S. Cassedy, "Waves guided by a boundary with time-space periodic modulation," \textit{Electr. Eng. Proc. Inst.}, vol. 112, no. 2 pp. 269-279, 1965.

\bibitem{CST}
CST Studio Suite, Computer Simulation Technolog ag, https://www.cst.com/

\bibitem{casaletti2019}
M. Casaletti, "Guided Waves on Scalar and Tensorial Reactance Surfaces Modulated by Periodic Functions: A Circuital Approach," \textit{IEEE Access}, vol. 7, pp. 68823-68836, 2019.

\bibitem{balanis2016}
C. A. Balanis, \textit{Antenna Theory: Analysis and Design}, 3rd ed. Hoboken,NJ, USA: Wiley, 2005.

\bibitem{martini2015}
E. Martini, M. Mencagli, and S. Maci, "Metasurface transformation for surface wave control," \textit{Philos. Trans. R. Soc. A}, vol. 373, no. 2049, p. 20140355, 2015.



\end{thebibliography}
\end{document}